\documentclass[12pt]{iopart}
  
\usepackage{graphicx}
\usepackage{dcolumn}
\usepackage{bm}
\usepackage{bbm}

\newcommand{\ave}[1]{\langle #1 \rangle}
\newcommand{\ve}[1]{\mathbf{#1}}
\newcommand{\bra}[1]{\langle #1|}
\newcommand{\ket}[1]{|#1\rangle}
\newcommand{\braket}[2]{\langle #1|#2\rangle}
\newcommand{\bracket}[3]{\langle #1|#2|#3 \rangle}
\def\ii{{\rm i}}

\usepackage{color}

\begin{document}

\title{Decoherence in regular systems}
\author{Marko \v Znidari\v c\dag\ddag and Toma\v z Prosen\dag}
\address{\dag\ Physics Department, Faculty of Mathematics and Physics, 
University of Ljubljana, Ljubljana, Slovenia}
\address{\ddag Department of Quantum Physics, University of Ulm, D-89069 Ulm, Germany}

\begin{abstract}
We consider unitary evolution of finite bipartite quantum systems and study time 
dependence of purity for initial cat states -- coherent 
superpositions of Gaussian wave-packets. We derive explicit formula 
for purity in systems with nonzero time averaged coupling, 
a typical situation for systems where an uncoupled part of the Hamiltonian 
is Liouville integrable. Previous analytical studies have been limited to harmonic oscillator systems 
but with our theory we are able to derive analytical results for 
general integrable systems. Two regimes are clearly identified, at short 
times purity decays due to decoherence whereas at longer times it decays 
because of 
relaxation.
\end{abstract}


\date{\today}

\maketitle

\section{Introduction}

Quantum mechanics is a linear theory and as such a superposition of 
solutions is also an admissible solution. Appearance of such 
superpositions at the quantum level and at the same time their 
absence from the 
macroscopic classical world has troubled scientists from the very 
beginning, an example is the famous ``Schr\" odinger Cat''~\cite{Schrodinger:35}. 
Their absence is usually explained in terms of {\em decoherence} due to 
the coupling with external degrees of freedom, see~\cite{Zurek:91} for a 
review. In recent years the process of decoherence has actually been 
experimentally measured~\cite{experimentParis,experimentNIST}. 
In the present paper we will derive decay of 
purity (1 - linear entropy) 
for initial cat states in finite Hamiltonian systems where an 
{\em uncoupled} part of the Hamiltonian generates regular (integrable) 
dynamics. Such systems are 
common among theoretical models (e.g. Jaynes-Cummings, ion trap quantum computer etc.) and can within certain approximation be even realized in the experiments. Our results are relevant also 
for the decoherence of macroscopic superpositions in general systems if 
the decoherence is faster than any dynamical time-scale involved. 
Note however that a strict decoherence in mathematical sense, 
i.e. irreversible loss of coherence, is possible only in the 
thermodynamic limit. Still, ``for all practical purposes'' one can have 
decoherence also in a sufficiently large finite system where the actual act of 
reversal is close to impossible due to high sensitivity to perturbations. 
Almost all theoretical studies of decoherence start from a master equation for
the reduced density matrix, see e.g.~\cite{Milburn:85} for the case of 
harmonic oscillator. Derivation of such a master equation is possible
only for very simple systems, for instance for a harmonic oscillator coupled 
to an infinite heat bath consisting of harmonic oscillators~\cite{Caldeira:83}.
Our approach here is different. We do not use master equation but rather 
start from the first principles, i.e. from a Hamiltonian describing system 
coupled to a finite ``bath''. In the case
of regular uncoupled dynamics we are able to derive the decay of purity for 
initial cat states. We stress that the result applies to any integrable 
dynamics, not just e.g. harmonic oscillators, and that the coupling can be 
quite arbitrary. Thus we will describe phenomena that go beyond 
simple ``reversible decoherence'' discussed for instance for two 
coupled cavities~\cite{Raimond:97}. Furthermore, our results might shed 
new light on the occurrence of decoherence in infinite-dimensional systems, 
e.g. harmonic oscillator bath~\cite{Caldeira:83}, by taking the 
appropriate limits.

\section{Process of decoherence}

Let us write the Hamiltonian of the entire system as
\begin{equation}
H=H_0+\delta\,\cdot V ,\qquad H_0=H_{\rm c} \otimes \mathbbm{1}_{\rm e}+\mathbbm{1}_{\rm c} \otimes H_{\rm e},
\label{eq:H}
\end{equation}
where $H_0$ is an uncoupled part of the Hamiltonian and $V$ is the coupling, with $\delta$ being its dimensionless strength. We will use 
subscripts ``c'' and ``e'' to denote ``central'' subsystem and ``environment'' (``environment'' is used just as a label for the part of the 
system we will trace over, without any connotation on its properties). 
In the present paper we study purity decay for the initial product state 
of the form
\begin{equation}
\ket{\psi(0)}=\frac{1}{\sqrt{2}}\left( \ket{\psi_{\rm c1}(0)}+\ket{\psi_{\rm c2}(0)} \right) \otimes \ket{\psi_{\rm e}(0)},
\label{eq:ic}
\end{equation}
with all three states $\ket{\psi_{\rm c1,2}(0)}$ and $\ket{\psi_{\rm e}(0)}$ 
being localized Gaussian wave packets. We assume that the initial state of the central system is composed of widely separated packets 
(i.e. is a cat state), so that two composing states are nearly orthogonal, 
$\braket{\psi_{\rm c1}(0)}{\psi_{\rm c2}(0)} \approx 0$. However, manipulations with macroscopic superpositions are quite nontrivial. In experimental situations the coherence between widely separated packets (e.g. $\ket{-\alpha}+\ket{\alpha}$ for large boson parameter $\alpha$) might be washed out due to presence of losses.~\cite{Yurke:86}. 
Purity is defined as
\begin{equation}
I(t)=\tr_{\rm c}{\rho_{\rm c}^2(t)},\qquad \rho_{\rm c}(t)=\tr_{\rm e}{\ket{\psi(t)}\bra{\psi(t)}},
\label{eq:I}
\end{equation}
where $\ket{\psi(t)}=\exp{(-\ii H t/\hbar)}\ket{\psi(0)}$. For our cat state (\ref{eq:ic}) the initial reduced density matrix reads
\begin{equation}
\fl \rho_{\rm c}(0)=\frac{1}{2}( \ket{\psi_{\rm c1}(0)}\bra{\psi_{\rm c1}(0)}+\ket{\psi_{\rm c2}(0)}\bra{\psi_{\rm c2}(0)} + \ket{\psi_{\rm c1}(0)}\bra{\psi_{\rm c2}(0)} + \ket{\psi_{\rm c2}(0)}\bra{\psi_{\rm c1}(0)} ).
\label{eq:rho0}
\end{equation} 
It has two diagonal terms and two off-diagonal terms also called coherences. 
These off-diagonal matrix elements are characteristic for coherent 
quantum superpositions. The process 
of decoherence then causes the {\em decay} of off-diagonal matrix elements of 
$\rho_{\rm c}(t)$, so 
that after some characteristic decoherence time $\tau_{\rm dec}$ we end up with 
the reduced density matrix $\rho_{\rm c}^{\rm mix}(t)$ which is a 
statistical {\em mixture} of two diagonal terms only,
\begin{equation}
\rho_{\rm c}^{\rm mix}(\tau_{\rm dec})=\frac{1}{2} (\ket{\psi_{\rm c1}(\tau_{\rm dec})}\bra{\psi_{\rm c1}(\tau_{\rm dec})}+\ket{\psi_{\rm c2}(\tau_{\rm dec})}\bra{\psi_{\rm c2}(\tau_{\rm dec})}).
\label{eq:rhomix}
\end{equation}
The purity of this reduced density matrix is $I(\tau_{\rm dec})=1/2$. Whereas 
the initial density matrix $\rho_{\rm c}(0)$ ($I(0)=1$) has no classical interpretation, 
the decohered density matrix $\rho_{\rm c}^{\rm mix}(\tau_{\rm dec})$ has. 
The aim of this paper is 
to explicitly show decoherence, i.e. to derive the transition 
\begin{equation}
\rho_{\rm c}(0) \stackrel{\tau_{\rm dec}}{\longrightarrow} \rho_{\rm c}^{\rm mix}(\tau_{\rm dec}).
\label{eq:dec}
\end{equation}

We are going to do this for systems with a regular uncoupled part $H_0$ of the 
Hamiltonian. The progress of decoherence will be ``monitored'' by calculating 
purity (\ref{eq:I}) which will decay from initial value $1$ to $1/2$ after 
$\tau_{\rm dec}$. When speaking about the decay of off-diagonal matrix 
elements of $\rho_{\rm c}(t)$ we should be a little careful though as the notion of off-diagonal depends on the basis. In has been recently 
shown~\cite{PRA,entangl} that the purity in regular systems and 
for initial localized wave packets decays on a very long time 
scale $\tau_{\rm p}\sim 1/\delta$, which is independent of $\hbar$. 
Localized wave packets are therefore very long-lived 
states -- the pointer states (they get entangled very slowly), 
and this is a preferred basis which we have in mind when speaking about the off-diagonal matrix elements. We will explain the meaning of pointer states more in detail later, after we derive analytic expression for purity decay. 
Provided $\tau_{\rm p}$ is larger than $\tau_{\rm dec}$ we can 
assume that during the process of decoherence propagation of initial 
constituent states still results in approximately product states, i.e. 
denoting $\ket{\psi_{1,2}(t)}=\exp{(-\ii H t /\hbar)} \ket{\psi_{\rm c1,2}(0)}\otimes\ket{\psi_{\rm e}(0)}$, we have for $t<\tau_{\rm p}$, $\ket{\psi_{1,2}(t)} \approx \ket{\psi_{\rm c1,2}(t)} \otimes \ket{\psi_{\rm e1,2}(t)}$. Note that we do not assume that $\ket{\psi_{\rm c,e}(t)}$ can be obtained by propagation with $H_{\rm c,e}$ alone, in fact they can not be, because, for instance, the state 
of the environment $\ket{\psi_{\rm e1,2}(t)}$ {\em depends} on the state of the central system. The above product form of individual states also justifies 
the form of the decohered density matrix $\rho_{\rm c}^{\rm mix}$ (\ref{eq:rhomix}) and is an important ingredient for the self-consistent explanation of the process of decoherence (\ref{eq:dec}).
The condition $\tau_{\rm p} > \tau_{\rm dec}$ means that the decoherence time-scale $\tau_{\rm dec}$ we are interested in is shorter than the relaxation time-scale $\tau_{\rm p}$ on which individual pointer states relax to their equilibrium. From the result of this paper (\ref{eq:pur_long}) we will see that this condition is satisfied provided the separation between two packets is large enough and/or $\hbar$ is sufficiently small.  

\section{Purity decay}

Let us now calculate off-diagonal matrix elements of $\rho_{\rm c}(t)$. They are of the form
\begin{equation}
\tr_{\rm e}{\ket{\psi_1(t)}\bra{\psi_2(t)}} \approx \ket{\psi_{\rm c1}(t)}\bra{\psi_{\rm c2}(t)} \braket{\psi_{\rm e2}(t)}{\psi_{\rm e1}(t)}.
\label{eq:off}
\end{equation}
Assuming the states of the central system are still approximately orthogonal, $\braket{\psi_{\rm c1}(t)}{\psi_{\rm c2}(t)}\approx 0$, the purity is simply 
$I(t)=\frac{1}{2}( 1 + F_{\rm e}(t))$, with
\begin{equation}
F_{\rm e}(t)=|\braket{\psi_{\rm e2}(t)}{\psi_{\rm e1}(t)}|^2 = \tr_{\rm e}{\rho_{\rm e1}(t) \rho_{\rm e2}(t)}.
\label{eq:Fe}
\end{equation}
Two reduced density matrices of the environment are $\rho_{\rm e1,2}(t)=\tr_{\rm c}{\ket{\psi_{1,2}(t)}\bra{\psi_{1,2}(t)}}$. Quantity $F_{\rm e}(t)$ gives the size of off-diagonal matrix elements and its decay is an indicator of decoherence. 
It is an overlap on the environment subspace of two states
at time $t$ obtained by the same evolution from two different initial product
states. It is similar 
to the fidelity~\cite{Prosen:02JPA}, which is the overlap of two states 
obtained from the same initial condition under two different evolutions. In 
fact $F_{\rm e}(t)$ can be connected with a quantity called reduced fidelity~\cite{Znidaric:04}, i.e. the fidelity on the subspace, see also~\cite{Recent} for a connection between decoherence and fidelity. 
Here we will not use this analogy as we will calculate $F_{\rm e}(t)$ directly.  Calculating $F_{\rm e}(t)$ is easier in the interaction picture, given by $\ket{\psi_{1,2}^{\rm M}(t)}= M(t) \ket{\psi_{1,2}(0)}$, with $M(t)=\exp{(\ii H_0 t /\hbar)}\exp{(-\ii H t /\hbar)}$ being the so-called echo operator used extensively in the theory of 
fidelity decay. Since $\exp(-\ii H_0 t/\hbar)$ 
factorizes, using $\rho_{\rm e1,2}^{\rm M}(t)$ instead of $\rho_{\rm e1,2}(t)$ will not change $F_{\rm e}(t)$. 
For systems where there exists an averaging time $t_{\rm ave}$ after which a time average of the coupling in the interaction picture, $V(t)=\exp{(\ii H_0 t /\hbar)}V\exp{(-\ii H_0 t /\hbar)}$, converges,
\begin{equation}
\bar{V}=\lim_{T \to \infty}{\frac{1}{T} \int_0^T{\!\!\!dt\, V(t)}},
\label{eq:Vbar}
\end{equation}
we can further simplify the echo operator. Note that $t_{\rm ave}$ is a classical time if the classical limit exists and effective $\hbar$ is sufficiently small, i.e. $t_{\rm ave}$ does not depend neither on $\delta$ nor on $\hbar$. It is given by the classical correlation time of the coupling $V(t)$. Nontrivial $\bar{V}$ will typically occur in regular (integrable) systems. But note that only $H_0$ needs to be regular whereas the coupling $V$ (and therefore also $H$) can be arbitrary. For $\delta \ll 1$ one can show that the leading order expression 
(in $\delta$) for the echo operator is
\begin{equation}
M(t)={\rm e}^{-\ii \delta t \bar{V}/\hbar}.
\label{eq:Mvbar}
\end{equation}
For details see e.g.~\cite{entangl}. We proceed with a semiclassical 
evaluation of $F_{\rm e}(t)$. The average $\bar{V}$ is by construction a 
function of action variables only and therefore the semiclassical evaluation 
of $F_{\rm e}(t)$ is simplified. The calculation goes along the same line as 
for the evaluation of fidelity~\cite{Prosen:02JPA} and of the purity for 
individual coherent initial states~\cite{entangl}. A sum over quantum numbers 
is replaced with an integral over the action space and quantum operator 
$\bar{V}$ is replaced with its classical limit $\bar{v}(\ve{j})$, where 
$\ve{j}=(\ve{j}_{\rm c},\ve{j}_{\rm e})$ is a vector of actions having 
$d=d_{\rm c}+d_{\rm e}$ components if central system and environment have 
$d_{\rm c}$ and $d_{\rm e}$ degrees of freedom, respectively. 
Let us consider Gaussian packets $\braket{\ve{j}_{a}}{\psi_{a}} \propto
\exp{\left\{-(\ve{j}_{a}\!-\!\ve{j}_{a}^*)\!\cdot\!\Lambda_{a}(\ve{j}_{a}\!-\!\ve{j}_{a}^*)/2\hbar + \ii \mathbbm{\theta}^*_{a}\cdot\ve{j}_{a}/\hbar \right\}}$, centered at $(\ve{j}_{a}^*,\mathbbm{\theta}^*_{a})$ where $\Lambda_{a}$ is a positive squeezing matrix determining its shape. The semiclassical expression of the initial density 
$p_{a}(\ve{j}_{a}) = \bracket{\ve{j}_{a}}{\rho_{a}(0)}{\ve{j}_{a}}$ reads
\begin{equation}
p_{a}(\ve{j}_{a})=\left(\frac{\hbar}{\pi} \right)^{\!\!d_{a}/2}\!\!\!\!\!\!\! \sqrt{\det{\Lambda_{a}}} \exp{\{-(\ve{j}_{a}\!-\!\ve{j}_{a}^*)\!\cdot\!\Lambda_{a}(\ve{j}_{a}\!-\!\ve{j}_{a}^*)/\hbar \}}.
\label{eq:packet}
\end{equation}
Subscript ``$a$'' takes values ``c1'', ``c2'' and ``e'', for three 
initial packets constituting the initial state (\ref{eq:ic}). Writing shortly $p_{1,2}(\ve{j})=p_{\rm c1,2}(\ve{j}_{\rm c}) p_{\rm e}(\ve{j}_{\rm e})$, we have the expression for $F_{\rm e}(t)$,
\begin{eqnarray}
F_{\rm e}(t)=\hbar^{-2d} \int{\!\! d\ve{j}\, d\tilde{\ve{j}}\, \exp{\left(-\ii \frac{\delta t}{\hbar} \Phi\right)} p_{1}(\ve{j}) p_{2}(\tilde{\ve{j}}) }, \nonumber \\
\Phi=\bar{v}(\ve{j}_{\rm c},\ve{j}_{\rm e})-\bar{v}(\tilde{\ve{j}}_{\rm c},\ve{j}_{\rm e})+\bar{v}(\tilde{\ve{j}}_{\rm c},\tilde{\ve{j}}_{\rm e})-\bar{v}(\ve{j}_{\rm c},\tilde{\ve{j}}_{\rm e}).
\end{eqnarray}
Note that the above integral is essentially a classical average (see~\cite{Gong:03} for some results) over two densities corresponding to two states $\ket{\psi_{1,2}(0)}$ and is therefore a sort of cross-correlation function. We expand the phase around the position $\ve{j}^*_{\rm e}$ of the environmental state,
$\Phi \approx \lbrack \bar{v}'_{\rm e}(\ve{j}_{\rm c})-\bar{v}'_{\rm e}(\tilde{\ve{j}}_{\rm c})\rbrack (\ve{j}_{\rm e}-\ve{j}_{\rm e}^*)- \lbrack \bar{v}'_{\rm e}(\ve{j}_{\rm c})-\bar{v}'_{\rm e}(\tilde{\ve{j}}_{\rm c})\rbrack (\tilde{\ve{j}}_{\rm e}-\ve{j}_{\rm e}^*),
$
where $\bar{v}'_{\rm e}(\ve{j}_{\rm c})=\partial \bar{v}(\ve{j}_{\rm c},\ve{j}_{\rm e}^*)/\partial \ve{j}_{\rm e}$ is a vector of partial derivatives with respect to the environment, evaluated at the 
position of the environmental packet $\ve{j}_{\rm e}^*$. Integration over the 
action variables now gives
\begin{eqnarray}
F_{\rm e}(t)=\exp{\left(-t^2/\tau_{\rm dec}^2 \right)},\qquad 
\tau_{\rm dec}= \sqrt{\hbar /C_{\rm e}}/\delta  \nonumber \\
C_{\rm e}=\frac{1}{2}\left( \bar{v}'_{\rm e}(\ve{j}_{\rm c1}^*)-\bar{v}'_{\rm e}(\ve{j}_{\rm c2}^*) \right) \Lambda_{\rm e}^{-1} \left( \bar{v}'_{\rm e}(\ve{j}_{\rm c1}^*)-\bar{v}'_{\rm e}(\ve{j}_{\rm c2}^*) \right).
\label{eq:Fe(t)}
\end{eqnarray}
We can see, that the decay time $\tau_{\rm dec}$ for cat state is indeed smaller than the decay time for individual states $\tau_{\rm p}\sim 1/\delta$, provided $\hbar \ll C_{\rm e}$. In the linear response regime $C_{\rm e}$ is just the semiclassical expression for
\begin{equation}
\fl C_{\rm e}=\frac{1}{\hbar} ( \ave{[\bar{V}-\ave{\bar{V}}_{\rm e}]^2}_{1} + \ave{[\bar{V}-\ave{\bar{V}}_{\rm e}]^2}_{2} +2 \ave{\bar{V}}_1\ave{\bar{V}}_2 -\ave{\ave{\bar{V}}_{\rm c1}\ave{\bar{V}}_{\rm c2}}_{\rm e} - \ave{\ave{\bar{V}}_{\rm c2}\ave{\bar{V}}_{\rm c1}}_{\rm e} ),
\label{eq:Ce}
\end{equation}
where the subscripts denote with respect to which initial state the average 
is performed.
Plugging $F_{\rm e}(t)$ (\ref{eq:Fe(t)}) into the expression for purity $I(t)$ we get purity decay for times $t<\tau_{\rm p}$. We can actually calculate the
purity for longer times as well. Namely, after $\tau_{\rm dec}$ the reduced 
density matrix is a statistical mixture of states $\ket{\psi_{\rm c1}(t)}$ and $\ket{\psi_{\rm c2}(t)}$. The purity can then be written as the sum of purities for individual states as there are no quantum coherences present anymore. Therefore, a complete formula for purity 
valid also for longer times is
\begin{equation}
I(t)=\frac{1}{4} \left( I_1(t)+I_2(t) + 2 F_{\rm e}(t) \right),
\label{eq:pur_long}
\end{equation}
where $I_{1,2}(t)=\tr_{\rm c}{\rho_{\rm c1,2}^2(t)}$ are purities of individual states. $I_{1,2}(t)$ decay on a long time scale $\tau_{\rm p}$ and have been calculated in~\cite{entangl}. 
Here we will just list 
the result
\begin{equation}
I_{1,2}(t)=\frac{1}{\sqrt{\det{(\mathbbm{1}+(\delta t)^2 u_{1,2})}}},\quad u_{1,2}=\Lambda_{\rm c1,2}^{-1} \bar{v}''_{\rm ce} \Lambda_{\rm e}^{-1} \bar{v}''_{\rm ec},
\label{eq:I12}
\end{equation}
with a $d_{\rm c} \times d_{\rm e}$ dimensional matrix $\bar{v}''_{\rm ce}$ of second derivatives evaluated at the position $\ve{j}^*_1$, or $\ve{j}^*_2$,
of the initial state,  
$( \bar{v}''_{\rm ce})_{kl}=\partial^2 \bar{v}(\ve{j}_{1,2}^*)/\partial (\ve{j}_{\rm c})_k \partial(\ve{j}_{\rm e})_l$.
\par
The formula for $I(t)$ (\ref{eq:pur_long}) is our main result. 
Let us discuss it more in detail. First, there are two time scales. A short one on which $F_{\rm e}(t)$ decays and a long one on which $I_{1,2}(t)$ decay. On a short time scale $\tau_{\rm dec}$ 
purity drops to $I(\tau_{\rm dec})=1/2$, signaling decoherence. 
This decay of $I(t)$ has a Gaussian form (\ref{eq:Fe(t)}) and is generally 
faster the further apart the centers of two 
initial states, $\ket{\psi_{\rm c1}(0)}$ and $\ket{\psi_{\rm c2}(0)}$, are. Expanding $\bar{v}'_{\rm c}(\ve{j}_{\rm c2}^*)$ around the position $\ve{j}_{\rm c1}^*$ we indeed have $C_{\rm e}\approx 
\frac{1}{2}[\bar{v}''_{\rm ec}\cdot (\ve{j}_{\rm c2}^*-\ve{j}_{\rm c1}^*)] 
\Lambda_{\rm e}^{-1}[\bar{v}''_{\rm ec}\cdot (\ve{j}_{\rm c2}^*-\ve{j}_{\rm c1}^*)]$. If the coupling $\bar{v}$ is between all pairs of degrees of freedom we have $C_{\rm e} \propto d_{\rm c} d_{\rm e} (\ve{j}_{\rm c2}^*-\ve{j}_{\rm c1}^*)^2$. The decay time $\tau_{\rm dec}$ (\ref{eq:Fe(t)}) is therefore 
inversely proportional to the distance between the packets 
$|\ve{j}_{\rm c2}^*-\ve{j}_{\rm c1}^*|$ and to $\sqrt{d_{\rm e}}$. 
Similar scaling of decoherence time with the number of degrees of 
freedom has been recently experimentally  measured in a NMR 
context~\cite{Krojanski:04}. Note that the decoherence time decreases, i.e. we have $\hbar \ll C_{\rm e}$, 
also when we approach the semiclassical limit $\hbar \to 0$. As the 
effective $\hbar$ determines the dimensionality of the relevant Hilbert 
space, $\tau_{\rm dec}$ will be smaller the larger Hilbert space we have. 
It also decreases with the number of degrees of freedom $d_{\rm e}$ of the environment, meaning decoherence gets faster for the environment with more degrees of freedom. Taking the thermodynamic limit is not straightforward though. The fact that the decoherence time $\tau_{\rm dec}$ goes to zero if the number of degrees of 
freedom $d_{\rm e}$ of the environment is increased at the constant coupling 
$\delta$ is not a surprise. Indeed, if we have a coupling of the same strength 
with infinitely many (and infinitely fast) degrees of freedom, 
decoherence will occur instantly. To have a physical thermodynamical limit one has to take $d_{\rm e} \to \infty$ and at the same time decrease the coupling strength to higher modes or make the so-called ``ultraviolet cutoff'' as in e.g. Caldeira-Leggett model~\cite{Caldeira:83}. In such a case one would get a finite $\tau_{\rm dec}$. For self-consistent description using $I(t)$ (\ref{eq:pur_long}) one also needs $\tau_{\rm dec} \gg t_{\rm ave}$. Note also that decoherence has a Gaussian shape in contrast to master equation approach where the decay 
is exponential. Gaussian form of decoherence has been obtained for 
macroscopic superpositions in~\cite{Haake}.
\par
Of course, all this holds provided the average perturbation $\bar{v}$ is nonzero. Interesting suppression of decoherence might arise for $\bar{v}=0$, for which extreme stability of quantum fidelity has recently been found~\cite{freeze}.
After decoherence time $\tau_{\rm dec}$ we are left with a statistical mixture of states and further decay of purity $I(t)$, due to the decay of $I_{1,2}(t)$, is a consequence of {\em relaxation} to final state. This relaxation happens on a long time scale $\tau_{\rm p}$ given by the slower decaying $I_{1,2}(t)$, see~\cite{entangl} for details. 
In order to illustrate the decay of $I(t)$ we performed numerical simulations.

\section{Numerical example}

\begin{figure}
\centerline{\includegraphics[angle=-90,width=13cm]{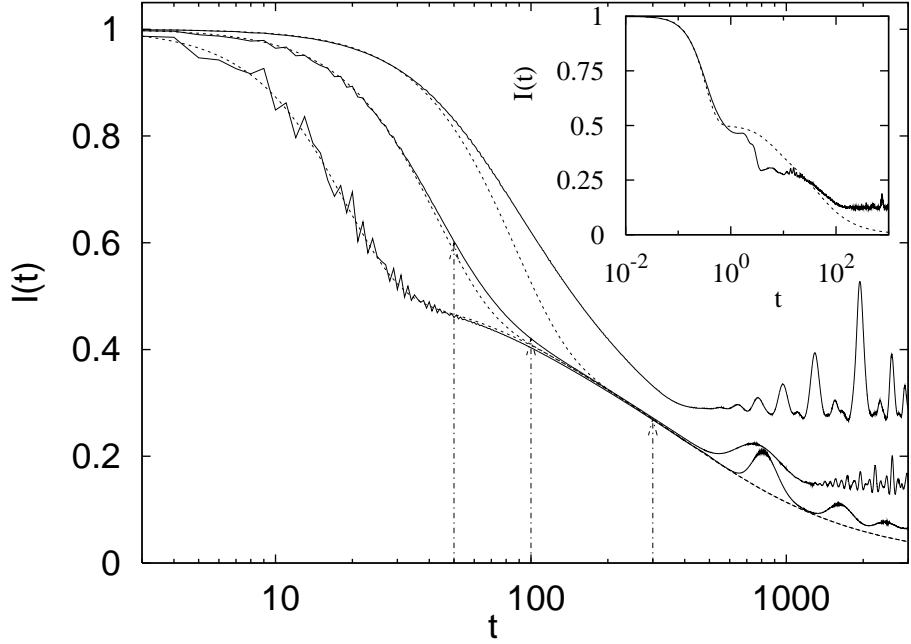}}
\caption{Purity decay for system (\ref{eq:H1x1}) for initial cat state and different $1/\hbar=25,100$ and $500$, full curves from right to left. Dotted curves converging to a single curve for $t>100$ are the theoretical prediction for $I(t)$ (\ref{eq:pur_long}) using decay times (\ref{eq:tdec},\ref{eq:I1x1}). Vertical arrows indicate times 
at which we show 
Wigner functions in Fig.~\ref{fig:wig}. In the inset we show the decay for macro-superposition, see text for details.}
\label{fig:catI}
\end{figure}

\begin{figure}[!ht]
\centerline{\includegraphics[width=15cm]{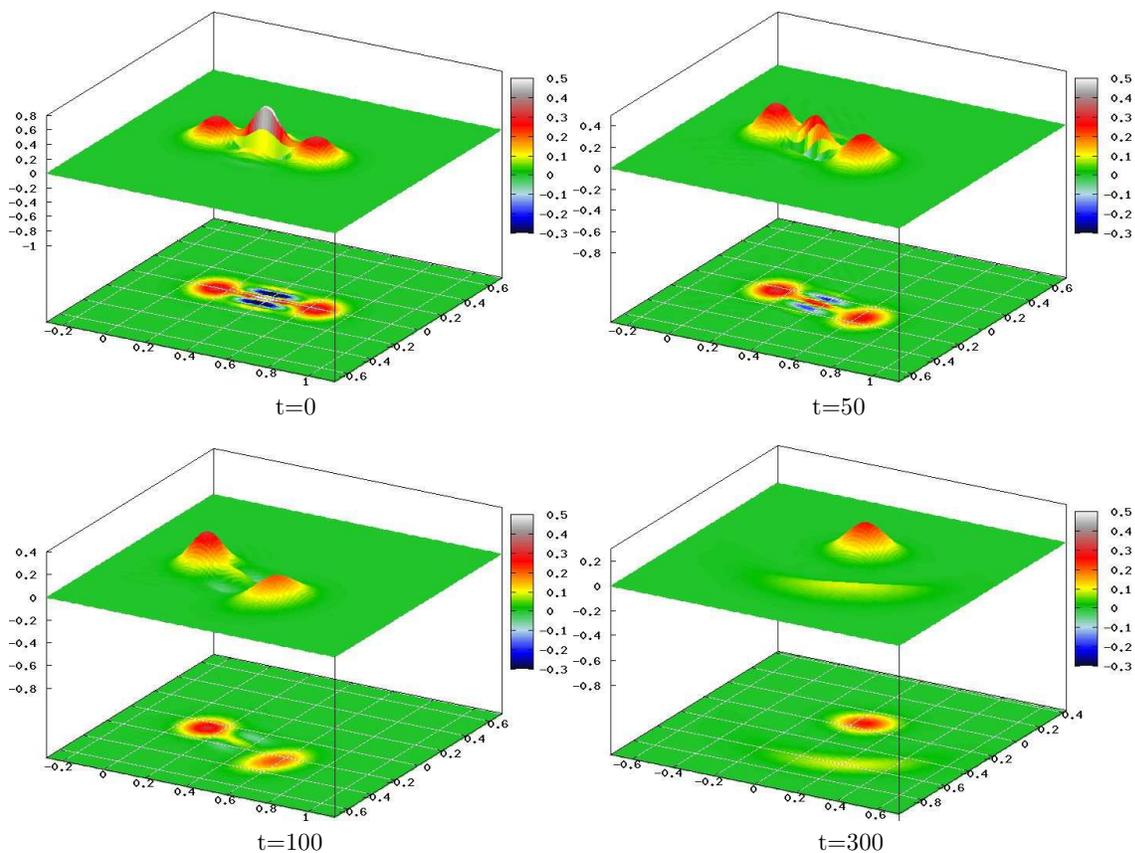}}
\caption{(Color online)Wigner functions for the reduced density matrix $\rho_{\rm c}^M(t)$ and initial cat 
state (\ref{eq:ic}). 
Planck constant is $\hbar=1/100$ all other data are the same as for 
Fig.~\ref{fig:catI}. On $x$ and $y$ axis we plot $\sqrt{2\hbar}\Re\alpha$ 
and $\sqrt{2\hbar}\Im\alpha$, respectively. Disappearance of the 
oscillations is an indicator of decoherence whereas the fading of 
packets is a signature of relaxation. See text for details.}
\label{fig:wig}
\end{figure}

We consider two coupled anharmonic oscillators,
i.e. $d_{\rm c}=d_{\rm e}=1$, with the Hamiltonian
\begin{equation}
H=\gamma_{\rm c} (\hbar a_{\rm c}^+ a_{\rm c}-\Delta)^2
+ \gamma_{\rm e} (\hbar a_{\rm e}^+ a_{\rm e}-\Delta)^2+ \delta \hbar^2 (a_{\rm c}^+ + a_{\rm c})^2(a_{\rm e}^+ + a_{\rm e})^2.
\label{eq:H1x1}
\end{equation}
where $a^+_{a},a_{a}$ denote boson raising/lowering operators. 
All initial wave packets are boson coherent states, 
$\ket{\psi_{\rm c,e}(0)}={\rm e}^{\alpha a^+ - \alpha^* a} \ket{0}$, where $\ket{0}$ is the ground state. The parameter $\alpha$ is chosen as $\alpha=\sqrt{j^*/\hbar}$ with the initial action $j_{\rm e}^*=0.1$ for the environment and $j_{\rm c1}^*=0.2$ and $j_{\rm c2}^*=0.01$ for the two states of the central system. The ``squeezing parameter'' $\Lambda$ (\ref{eq:packet}) is for coherent states equal to $\Lambda=1/(2j^*)$. Other parameters of the Hamiltonian are $\gamma_{\rm c}=1$, $\gamma_{\rm e}=0.6456$ and 
$\Delta=1.2$. Coupling strength is set to $\delta=0.01$, but note that our 
theory is often not limited to small $\delta$~\cite{entangl}. Time averaged coupling is easily calculated from the classical limit of the Hamiltonian and is
$\bar{v}=4 j_{\rm c} j_{\rm e}$. Using this we easily evaluate 
$C_{\rm e}$ (\ref{eq:Fe(t)}) 
\begin{equation}
C_{\rm e}=16 j_{\rm e}^* (j_{\rm c1}^*-j_{\rm c2}^*)^2\approx 0.058,
\label{eq:Cen}
\end{equation}
and the decay time $\tau_{\rm dec}$ of $F_{\rm e}(t)$ (\ref{eq:Fe(t)}), 
\begin{equation}
\tau_{\rm dec}=
\sqrt{\hbar/j_{\rm e}^*}/
(4\delta |j_{\rm c1}^*-j_{\rm c2}^*|) \approx 416 \sqrt{\hbar}.
\label{eq:tdec}
\end{equation}
The matrix $u$ needed in $I_{1,2}(t)$ (\ref{eq:I12}) is just a 
number equal to $u_{1,2}=64 j_{\rm e}^* j_{\rm c1,2}^*$. Theoretical decay of 
$I_{1,2}(t)$ (\ref{eq:I12}) is therefore
\begin{equation}
I_{1,2}(t)=\frac{1}{\sqrt{1+(t/\tau_{\rm p1,2})^2}},\quad \tau_{\rm p1,2}=\frac{1}{8\delta\sqrt{j_{\rm c1,2}^* j_{\rm e}^*}}.
\label{eq:I1x1}
\end{equation}
In Fig.~\ref{fig:catI} we show the results of numerical simulation for three different values of $\hbar$. For the smaller values of $\hbar$ we can clearly see the two regimes discussed. Initially $I(t)$ decays due to decoherence as described by $F_{\rm e}(t)$. After the decoherence time $\tau_{\rm dec}$, i.e. after $I(t)$ falls to the value $1/2$, relaxation begins. This regime is described by two decaying purities $I_{1,2}(t)$ of individual states and does not depend on $\hbar$. 

For the largest value of $\hbar=1/25$ finite size effects can be observed.
The origin of finite size effects is twofold: First, the condition of well separated decoherence time scale $\tau_{\rm dec}$ ($\hbar$-dependent) and relaxation time scale $\tau_{\rm p}$ ($\hbar$-independent) results in $\hbar \ll C_{\rm e}$. Second, there is a finite saturation value of purity for nonvanishing $\hbar$, $I(t \to \infty) \sim \hbar^{d_{\rm c}}$. For the largest $\hbar=1/25$ shown in Fig.~\ref{fig:catI}, the saturation value is $I\sim 0.3$ which is almost as large as purity at the end of decoherence, $I(\tau_{\rm dec})=1/2$. Therefore this finite saturation value will start to influence purity decay already at $\tau_{\rm dec}\approx 80$ and no relaxation can be observed. 

In Fig.~\ref{fig:wig} we show a Wigner function of the reduced density matrix $\rho_{\rm c}^{\rm M}(t)=\tr_{\rm e}{M(t)\rho(0)M^\dagger(t)}$ at times $t=0,50,100$ and $t=300$. The integral of the square of the Wigner function equals purity. At $t=0$ we see strong oscillations in the Wigner function at the midpoint between two packets. This is a characteristic feature of coherent superpositions. At $t=50$ we are in the middle of decoherence where the oscillations and negative values of the Wigner function have already been reduced. At $t=100$ the process of decoherence has ended, visible as the absence of negative regions in the Wigner function. At still longer time $t=300$ the relaxation is in full swing as the second packet has almost entirely disappeared. In Fig.~\ref{fig:wig} one can observe that the packets rotate around the origin and that the relaxation rates of the two packets are not equal. This can be explained by the form of the time averaged Hamiltonian in the interaction (i.e. echo) picture, $\bar{v}=4 j_{\rm c}j_{\rm e}$. This Hamiltonian causes rotation of the angle of the central system, $\dot{\theta}_{\rm c}=4 j_{\rm e}$. Using the value of $j_{\rm e}^*=0.1$ we get that the packets should rotate for $90^0$ around the origin in time $t\approx 400$, which agrees with the data shown. Unequal relaxation can be simply understood from the expression for $\tau_{\rm p1,2}$ (\ref{eq:I1x1}). The packet with smaller $j_{\rm c}^*$ will relax on a longer time scale. This slower decay is a second order effect due to smaller size of the packet in the action direction (i.e. larger $\Lambda$) which causes a slower dephasing of angles $\theta_{\rm e}$ and in turn slower decay of purity $I(t)$.   

We should mention that at very large times (i.e. much later after the relaxation) we get revivals of purity. These are a simple consequence of having a finite number of eigenmodes of the uncoupled system for not too large $\hbar$, which causes a complex beating-like phenomena at large times. Note that these revivals are not simple Rabi oscillations between the two subsystems as discussed for instance for cavities in~\cite{Raimond:97}. For our data (not shown in the figure) and times $t<10000$ we have a revival $I\approx 0.95$ at $t\approx 4000$ for $\hbar=1/25$, revival of $I\approx 0.5$ at $t\approx 8000$ for $\hbar=1/100$ and there is no revival for $\hbar=1/500$ (and $t<5000$). The revivals are therefore less prominent and happen at larger times for small $\hbar$. The same would happen if the central system would have more than one degree of freedom. For many degrees of freedom systems, high sensitivity to perturbations will also effectively prevent the possibility of restoring coherence upon time-reversal.

Finally let us comment on the relevance of our results for the appearance 
of 
the macro world. It has been pointed out~\cite{Haake} that the 
decoherence time for 
sufficiently macroscopic superpositions gets very short. Therefore, the 
system's dynamics can be considered to be regular on this very short 
time scale, but on the other hand we need $\tau_{\rm dec}\gg t_{\rm ave}$ 
for our theory to apply ($t_{\rm ave}$ does not depend on $\delta$ or $\hbar$). Resolution of this problem is immediate if one 
looks at the quantum expression for $C_{\rm e}$ (\ref{eq:Ce}). For short times one has to replace $\bar{V}$ with an ``instantaneous'' operator $V(0)$ and 
evaluate 
quantum expectation values using classical averaging. Doing this for our 
model (\ref{eq:H1x1}) for instance, we get 
$\tau_{\rm dec}(t \ll t_{\rm ave})=\tau_{\rm dec}/4$, 
with $\tau_{\rm dec}$ given in (\ref{eq:tdec}). This theoretical prediction 
agrees with the numerical curve for $\delta=0.06$, 
$\hbar=1/100$ and $j_{\rm c2}^*=1$, shown in the inset of Fig.~\ref{fig:catI}, where $\tau_{\rm dec}$ is much shorter than for the three curves in the 
main plot. Again, the theoretical curve (dotted) agrees with the numerics (full line). Our theory therefore covers regime $\tau_{\rm dec} \gg t_{\rm ave}$ as well as $\tau_{\rm dec} \ll t_{\rm ave}$.

\section{Conclusions}
We have derived an analytic semi-classical expression for the purity decay 
of initial cat states in finite composite systems which are integrable in the absence of coupling and have a nonzero time averaged coupling. Note that the coupling can in general break integrability. Such systems are found both among 
theoretical models and in experiments. Pointer states are identified as localized action packets in the action components for which the time averaged coupling in nontrivial. Purity for superpositions of such pointer states first decays  on a short time scale as a Gaussian, indicating decoherence. On longer time scale, after the decoherence, it decays in an algebraic way, signifying relaxation to equilibrium. Theoretical results are confirmed by the numerical simulation of two coupled nonlinear oscillators.

\section*{Acknowledgments}

We thank Thomas H. Seligman for fruitful discussions,
and CiC (Cuernavaca, Mexico), 
where parts of this work have been completed, for hospitality.
The work has been financially supported by the grant P1-0044 of the Ministry 
of higher education, science and technology of Slovenia, and in part by the 
ARO grant (USA) DAAD 19-02-1-0086. 
M\v Z would like to thank AvH Foundation for the support during the last 
stage of the work. We would also like to thank anonymous referee for pointing out the reference~\cite{Yurke:86}.

\end{document}